\documentclass[11pt, letterpaper]{article}

\usepackage[letterpaper,left=1in,right=1in,top=1in,bottom=1in]{geometry}
\usepackage{graphicx}
\usepackage{amsmath}
\usepackage{amssymb}
\usepackage{setspace}
\usepackage{epstopdf}
\usepackage{color}
\usepackage{algorithm}
\usepackage{algorithmicx}
\usepackage{algpseudocode}
\usepackage{todonotes}
\usepackage{booktabs}
\usepackage{multirow}
\usepackage{cite}
\usepackage{siunitx,etoolbox} % table font bond
\usepackage{arydshln}
\usepackage{subfigure}
\usepackage[normalem]{ulem}

\newtheorem{rmk}{Remark}
\usepackage{url}
\title{\Large \bf Machine learning-based input-augmented Koopman modeling and predictive control of nonlinear processes}

\author{
\centerline{\normalsize Zhaoyang Li$^{a}$, Minghao Han$^{a,b}$, Dat-Nguyen Vo$^{a}$, Xunyuan Yin$^{a,b,}$\thanks{Corresponding author: X. Yin. Tel: (+65) 6316 8746. Email: xunyuan.yin@ntu.edu.sg.}}
\vspace{5mm}\\
\centerline{\small $^{a}$School of Chemistry, Chemical Engineering and Biotechnology, Nanyang Technological University,}\\
\centerline{\small 62 Nanyang Drive, 637459, Singapore}\\
\centerline{\small $^{b}$ Environmental Process Modelling Centre, Nanyang Environment and Water Research Institute (NEWRI),}\\
\centerline{\small Nanyang Technological University, 1 CleanTech Loop, 637141, Singapore}\\
}
\allowdisplaybreaks

\date{}

\begin{document}

\maketitle
\thispagestyle{plain}
\pagestyle{plain}
\setstretch{1.5}

\begin{abstract}

Koopman-based modeling and model predictive control have been a promising alternative for optimal control of nonlinear processes. Good Koopman modeling performance significantly depends on an appropriate nonlinear mapping from the original state-space to a lifted state space. In this work, we propose an input-augmented Koopman modeling and model predictive control approach. Both the states and the known inputs are lifted using two deep neural networks (DNNs), and a Koopman model with nonlinearity in inputs is trained within the higher-dimensional state-space. A Koopman-based model predictive control problem is formulated. To bypass non-convex optimization induced by the nonlinearity in the Koopman model, we further present an iterative implementation algorithm, which approximates the optimal control input via solving a convex optimization problem iteratively. The proposed method is applied to a chemical process and a biological water treatment process via simulations. The efficacy and advantages of the proposed modeling and control approach are demonstrated.

\end{abstract}

\section{Introduction}

Advanced process control is crucial for managing the real-time operation of modern industrial processes, ensuring safe operation and efficient production, and reducing energy consumption and waste\cite{daoutidis2018integrating,christofides2013distributed,daoutidis2016sustainability}. Model predictive control (MPC) has been the most commonly used advanced process control solution for nonlinear industrial processes. 
A Lyapunov-based MPC was proposed in \cite{heidarinejad2012economic} to optimize the economic performance of the nonlinear system while ensuring the states bounded in the predefined stability region.
In \cite{ferramosca2010mpc}, zone MPC was proposed to deal with the control objectives in the form of a target set; this method largely increased the degree of freedom of the control.
In \cite{lee1994extended}, extended Kalman filtering was integrated into MPC for output-feedback control.
In \cite{zhang2013lyapunov}, a moving horizon estimator was utilized to estimate the full states of nonlinear processes. Based on the estimated states, an MPC with guaranteed stability was developed.
More results on nonlinear MPC can be found in \cite{kummer2020semi,palma2022integration}.

The efficacy and performance of MPC are dependent on the open-loop predictability of the adopted dynamic model within the control/prediction horizon. In the context of nonlinear MPC, most of the existing methods, including the aforementioned designs, have been developed based on the premise that accurate first-principles models are available \cite{han2013nonlinear,zhang2019economic}. However, obtaining first-principles knowledge can be expensive or even impossible for complex processes with complex kinetics and mass and energy balances. This has substantially limited the applications of these conventional nonlinear MPC approaches. 
Data-driven modeling holds the promise to develop MPC for nonlinear processes without the need for an accurate physics-based dynamic model\cite{rajulapati2022integration}. 
For example, a formulation of a data-driven model based on the Hankel matrix was proposed in \cite{berberich2020data}. A linear model can be established via the collected input-output trajectory, without requiring any physical knowledge.
Machine learning methods including deep neural networks (DNNs) have been applied to characterize the intricate dynamic behaviors of various nonlinear processes \cite{wu2019machine}.
In \cite{zheng2022online}, 
recurrent neural networks (RNNs) were exploited to describe the batch crystallization process including time series data.  {\color{black}In addition, hybrid modeling approaches that integrate the available physical knowledge and data information have been proposed to describe the dynamic behaviors of nonlinear processes for enhanced model robustness \cite{bangi2020deep,pahari2024achieving}.}

The Koopman theory \cite{koopman1931hamiltonian,wynn2013optimal} holds promise for modeling and controlling nonlinear processes based on data. By characterizing nonlinear dynamics in a linear manner, convex MPC can be developed and implemented despite the nonlinearity of the process dynamics; this makes Koopman-based MPC approaches more computationally efficient and applicable than those based on DNNs\cite{otto2021koopman}. In \cite{schulze2022identification}, the Koopman operator for autonomous systems was generalized to non-autonomous systems, which enables the use of Koopman operators for describing the dynamics of general controlled nonlinear systems.
Based on the Koopman operator, linear convex MPC can be developed for nonlinear systems \cite{schulze2022identification,korda2018linear}.
From an application standpoint, a significant challenge in Koopman modeling lies in identifying the observable functions that establish the nonlinear mapping between the original nonlinear state-space and a lifted linear state-space, where an approximate Koopman operator linearly governs the nonlinear dynamics\cite{brunton2022modern}. 
The extended dynamic mode decomposition(EDMD) \cite{williams2015data} has been a representative method for Koopman operator approximation; this method exploits manually selected observable functions to account for the nonlinear mapping and establish Koopman operators via solving a least-squares problem. Based on EDMD, a sparse identification of nonlinear dynamics (SINDy)-based Koopman modeling and control approach was proposed \cite{zhang2023reduced}. In this approach, the most appropriate observable functions were selected from a pre-determined rich library through recursive Kalman filtering. {\color{black}Using EDMD, an offset-free Koopman MPC method was proposed in \cite{son2022development}; the stability of a Lyapunov-based Koopman MPC design was analyzed in \cite{narasingam2023data}}.
An alternative solution to Koopman modeling and control is to train neural networks to represent the observable functions \cite{han2020deep,yeung2019learning}.
In \cite{han2023robust}, RNNs were leveraged for learning observable functions in Koopman modeling and control of time-delay nonlinear processes.
Learning-based Koopman modeling can bypass the need for manually selecting appropriate observable functions, a task that can be exhaustive and challenging as the size and scale of the considered process increase.

 {\color{black}We have witnessed the applications of Koopman-based MPC in various fields, including robotics \cite{wang2022improved}, biological systems \cite{Koopman_empc}, chemical processes\cite{zhang2023reduced,son2021application,narasingam2019koopman, CEP2021} and vehicular systems \cite{chen2024incorporating}.
%  In \cite{narasingam2019koopman},
% Koopman-based MPC is applied on a continuously stirred tank reactor (CSTR) system. For the pulping process, a Koopman-based MPC is designed to produce
% pulp with desired properties under the influence of feed fluctuations\cite{CEP2021}. 
}
Meanwhile, it is worth pointing out that in
the above-mentioned Koopman modeling and MPC methods, including the learning-enabled methods, typically only the original states are projected to a lifted space, with the
hypothesis that the states in the lifted higher-dimensional state space are linearly dependent
on the inputs of the underlying nonlinear system. However, this hypothesis and treatment may result in an inadequate characterization of the nonlinear dynamics of the underlying process. One exemption is the method proposed in \cite{shi2022deep}. Both the inputs were transformed into a lifted space via neural networks. Therefore, a nonlinear Koopman model was established. To preserve the convexity of the associated optimal control problem, the encoded inputs in the lifted space were treated as the decision variables, and an inverse mapping was established to recover the actual control action that needs to be applied to the physical system. However, finding the inverse mapping can be challenging. Additionally, based on this Koopman modeling approach, it has been difficult to address hard constraints on the control inputs.

In this work, we revisit machine learning-based Koopman modeling and MPC for general nonlinear processes. We acknowledge and explicitly take into account the potential nonlinear dependence of the state variables in a lifted higher-dimensional space on known inputs of the original nonlinear process. Accordingly, two DNNs are exploited to characterize the nonlinear dependence of the lifted state on both the states and known inputs in the original nonlinear state space. A nonlinear Koopman model is established based on the proposed modeling approach. An iterative MPC implementation method is proposed to preserve the convexity of Koopman MPC. The proposed method is applied to a benchmark chemical process and a biological water treatment process. The advantages of the proposed method in terms of both modeling and control are illustrated via comparisons to the existing representative methods. The contributions of this work include:
\begin{enumerate}
    \item An input-augmented learning-based Koopman modeling approach is proposed to account for the nonlinear dependence of the lifted state variables on both the states and known inputs of the underlying nonlinear system for enhanced multi-step-ahead predictability of a Koopman model.
    \item An iterative MPC implementation method is developed to maintain the convexity of the optimization problem associated with Koopman MPC despite the nonlinearity of the established Koopman model.  
    \item  Simulations are conducted on a chemical process and a biological process to illustrate the efficacy and superiorities of the proposed modeling and control approach.
\end{enumerate}

\section{Preliminaries and problem formulation}

\subsection{Koopman operator}
Consider a general autonomous nonlinear system in the following form:
\begin{equation}
    x_{k+1}=f(x_k)
    \label{eq1}
\end{equation}
According to the Koopman theory\cite{koopman1931hamiltonian}, there exists an infinite-dimensional Koopman operator $\mathcal{K}: \mathcal{H}\rightarrow\mathcal{H}$ that acts on functions of the original state space to linearly governs the dynamics of the nonlinear processes in \eqref{eq1} as follows: 
\begin{equation}
\mathcal{K}\psi(x_k)=\psi \circ f(x_k) = \psi(x_{k+1})
\end{equation}
where $\psi$ represents the observable functions on the lifted space $\mathcal{H}$, and $\circ$ denotes function composition. 

Exploring the exact infinite-dimensional Koopman operator can be challenging for real-world nonlinear systems. From a practical perspective, constructing finite-dimensional approximations of Koopman operators is more feasible and ensures computational traceability. This involves creating a finite-dimensional function space $\overline{\mathcal{H}}\subset\mathcal{H}$ using a set of linearly independent functions.

\subsection{Koopman modeling for controlled systems}

The Koopman operator theory provides a promising framework for handling nonlinear control problems in a linear and efficient manner. Various methods have been proposed to extend the capability of the Koopman operator to describe controlled nonlinear systems \cite{korda2018linear,zhang2022robust}. 
The controlled nonlinear systems can be presented in the following form:
\begin{equation}
x_{k+1}=f(x_k,u_k)
\label{eq_u_k}
\end{equation}
where $x_{k}\in \mathcal{X} \subset\mathbb{R}^n$ is the state vector at time instant $k$; $u_{k}\in \mathcal{U} \subset \mathbb{R}^{m}$ denotes the input vector of the system at time instant $k$; $f$ is a nonlinear function describing the dynamic behavior of the nonlinear system. 

Following the approach proposed in \cite{korda2018linear,zhang2022robust}, an augmented state vector is constructed as $\mathcal{X}_k = \left[x_k^{\text{T}}, u_k^{\text{T}}\right]^{\text{T}}$ to include both the original states and inputs.
The dynamics of $\mathcal{X}_k$ can be described as 
\begin{equation}
\label{chi}
\mathcal{X}_{k+1}=\mathcal{F}(\mathcal{X}_{k}):=\begin{bmatrix}
                    f(x_k,u_k)\\
                    \mathcal{S}u_k\\
                    \end{bmatrix}:=\begin{bmatrix}
                    f(x_k,u_k)\\
                    u_{k+1}\\
                    \end{bmatrix}
\end{equation}
where $\mathcal{S}$ is a left shift operator, defined as $\mathcal{S}u_k=u_{k+1}$.

The Koopman operator for \eqref{chi}, denoted by $\mathcal{K}_\mathcal{F}:\mathcal{H}_\mathcal{F}\rightarrow\mathcal{H}_\mathcal{F}$, linearly governs the dynamics of the augmented vector $\mathcal{X}$ as follows:
\begin{equation}
\mathcal{K}_\mathcal{F}\Psi(\mathcal{X}_k) = \Psi \circ \mathcal{F}(\mathcal{X}_k) =  \Psi(\mathcal{X}_{k+1}) .
\label{koopman_1}
\end{equation}
where $\Psi$ contains the observable functions that map the original states to the lifted linear state space.
% Approaches including EDMD\cite{williams2015data} and neural networks\cite{yeung2019learning} can be applied to obtain the suitable observable function. 

The states and the inputs involved in $\mathcal{X}$ are treated separately in \cite{korda2018linear} 
when specifying the finite-dimensional observable functions. Particularly, the observable functions $\Psi$ are determined as: 
\begin{equation}\label{orignial form}
\Psi (\mathcal{X}_k) =\Psi (x_k,u_k) = \left[\psi^{\text{T}}(x_k),\mathcal{L}^{\text{T}}(u_k)\right]^{\text{T}}
\end{equation}
where $\psi:\mathbb{R}^{n}\rightarrow\mathbb{R}^{N}$  and $\mathcal{L}:\mathbb{R}^{m}\rightarrow\mathbb{R}^{M}$ are functions that transform the original states and inputs, respectively.

Since the objective is to use the Koopman operator to predict the future states but not to predict the future inputs,
we only need to reconstruct the elements in the first $N$ rows of the Koopman operator. Specifically, let ${\mathcal K}_{N_{\phi}}$ denote a finite-dimensional approximation of the Koopman operator. This operator can be represented by the following block matrices:
\begin{equation}\label{koopman operator}
{\mathcal K_{N_\phi}} = \left[
    \begin{array}{c;{2pt/2pt}c}
        A&B \\ \hdashline[2pt/2pt]
        *&*
    \end{array}
\right]
\end{equation}\normalsize
where $A\in\mathbb{R}^{N\times N}$, $B\in\mathbb{R}^{N\times M}$. In \eqref{koopman operator}, only $A$ and $B$ need to be identified, while the blocks represented by $*$ can be neglected. 

In \cite{korda2018linear}, $\mathcal{L}$ is determined as $\mathcal{L}(u_k)=u_k$ to ensure the linearity of the resulting Koopman model with respect to the control inputs. Consequently, a Koopman model is formulated as:
\begin{equation}
    \begin{aligned}
    \label{eq2}
    z_{k+1}&=Az_k+Bu_k\\
    \hat{x}_{k}&= Cz_{k}\\
    % z_k &= \psi(x_k)\\
    \end{aligned}
\end{equation}
where $z_k= \psi(x_k)$
% \in\mathbb{R}^{N}$
represents the state vector in the lifted space $\overline{\mathcal{H}}$, and $C\in\mathbb{R}^{n\times N}$ is a matrix that maps the states from space  $\overline{\mathcal{H}}$ to original space.
This Koopman model can be used to predict the future behavior of the nonlinear system in (\ref{eq_u_k}) in the lifted state space $\overline{\mathcal{H}}$. 
With the Koopman model in (\ref{eq2}), linear control approaches can be employed to govern the nonlinear system in (\ref{eq_u_k}).

\subsection{Observations and problem formulation}
% The control performance is largely dependent on the accuracy of prediction. 
% As the prediction result $\hat{x}_{k+1}$ converges toward $x_{k+1}$, the output of the corresponding controller becomes increasingly close to the optimal controller for nonlinear system $(\ref{eq1})$. Multiple methodologies are proposed to reduce the residual of prediction \cite{brunton2016sparse,liu2024koopa} (especially for long-term prediction). 
% % The choice and structure of the observable function $\boldsymbol{\psi}$ are widely discussed in \cite{brunton2016sparse,liu2024koopa}, 
% However, the predictors remain in the form of (\ref{eq2}) in most studies.

The method in \cite{korda2018linear} presents a promising framework for establishing linear Koopman models suitable for a wide range of controlled nonlinear systems. Meanwhile, due to the specific choice of $\mathcal{L}(u_k)=u_k$, this method does not account for the inherent nonlinear dependence of the lifted states on the known system inputs, as indicated by $\Psi(x_k,u_k)$ in (\ref{orignial form}). 
Treating this nonlinear dependence on the input in a linear manner may lead to compromised modeling performance, as will be demonstrated in Section \ref{experiment1} and Section \ref{experiment2} of this paper. 
In this work, we aim to address this gap while ensuring the formulated MPC based on the established Koopman model can be solved via quadratic programming.

% Detailed information about the predictor and corresponding controller is presented in the next sections.

\section{Learning-Based Koopman modeling with input augmentation}\label{sec:model}

% \todo{rewrite this}
% In this section, we present the proposed learning-based Koopman modeling approach with input augmentation. {\color{red}This method allows for the construction of a Koopman model for the nonlinear process with known input. The formulation of the process can be presented as 
In this section, we consider general nonlinear processes with both control inputs and uncontrolled inputs (known disturbances) as follows:
\begin{equation}\label{paper:1:nonlinear:p}
x_{k+1}=f(x_k,u_k,p_k)
\end{equation}
where $p_k\in \mathbb{R}^{p}$ denotes known disturbances to the system at time instant $k$.

\begin{figure}[t]
  \centering
\includegraphics[width=0.9\textwidth]{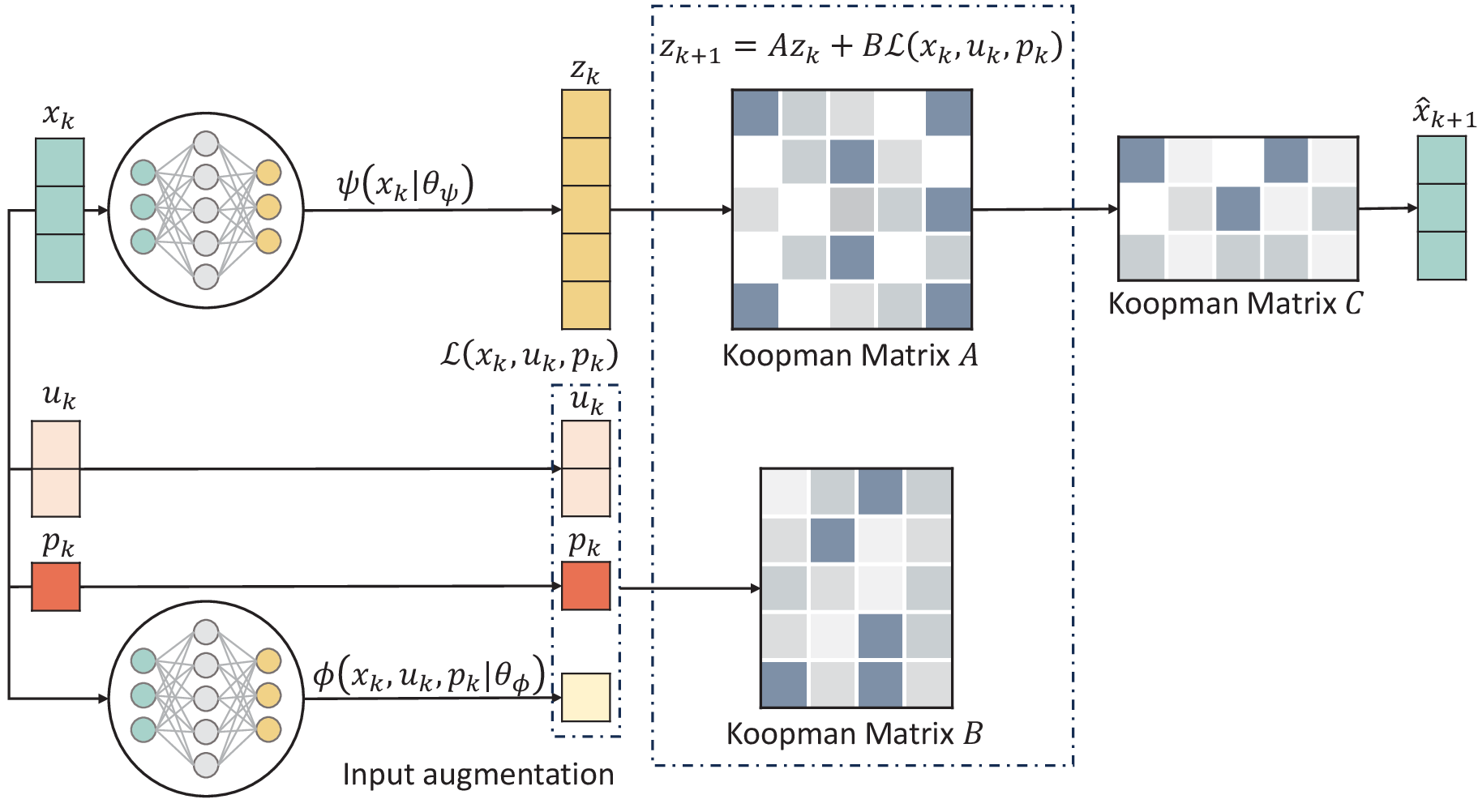}
  \caption{A graphic representation of the proposed Koopman modeling approach with input augmentation.}  
  \label{figure:1}
\end{figure}

{\color{black}
As shown in \eqref{paper:1:nonlinear:p}, the future states of the considered process can have nonlinear dependence on both the current states and inputs.} Accordingly, we propose a modeling approach named Deep Koopman Operator with input augmentation (DKOIA) for \eqref{paper:1:nonlinear:p}. An overview of the proposed approach is shown in Figure~\ref{figure:1}. In this approach, two DNNs are employed to account for the nonlinear mappings from the original states and known inputs to the lifted higher-dimensional space, in which the state vector is represented by $z$. Through characterizing the evolution of $z$ using the Koopman matrices $A$ and $B$, the dynamic behaviors of the original nonlinear system in \eqref{paper:1:nonlinear:p} are described. A prediction of the original state vector, denoted by $\hat x$, is obtained through a linear projection using matrix $C$. 

\subsection{Koopman model with input augmentation}
Considering the known inputs $u_k$ and $p_k$, an augmented state vector is constructed as  $\mathcal{X}_k = \left[x_k^{\text{T}}, u_k^{\text{T}},p_k^{\text{T}}\right]^{\text{T}}$. Instead of using ad hoc $\mathcal L = ([u^{\text{T}}, p^{\text{T}}]^{\text{T}})$ from \cite{korda2018linear}, a nonlinear $\mathcal L (x_k,u_k,p_k)$ is designed to transform both the states and inputs. Specifically, $\mathcal L (x_k,u_k,p_k) := \left[u_k^{\text{T}},p_k^{\text{T}},\phi^{\text{T}}(x_k,u_k,p_k)\right]^{\text{T}}$, where $\mathcal L : \mathbb{R}^m\times\mathbb{R}^n\times \mathbb{R}^p\rightarrow\mathbb{R}^{M+m+p}$ and $\phi : \mathbb{R}^m\times\mathbb{R}^n\times \mathbb{R}^p\rightarrow\mathbb{R}^{M}$. 

The usage of these observable functions leads to a Koopman model in the following form:
\begin{equation}
    \psi(x_{k+1}) = A \psi(x_k) +  B \mathcal L (x_k, u_k, p_k)\label{paper1:form}
\end{equation}
where $A$ and $B$ are the matrices that will be constructed using data.

By using $z$ to represent the lifted state vector, that is, $z_k:= \psi(x_k)$, the Koopman model with input augmentation is presented in the following form:
\begin{subequations}    \label{eq3}
    \begin{align}
    \label{eq:state-space}
    z_{k+1}&=Az_k+B \mathcal L (x_k, u_k, p_k)\\
    \label{eq:trans_back}
    \hat{x}_{k} &= Cz_{k}
    \end{align}
\end{subequations}

\subsection{Neural network-based observable functions}

Neural networks are applied to learn the observable functions from data. 
In this work, 
nonlinear functions $\psi$ and $\phi$ are represented by neural networks with trainable parameters $\theta_\psi$ and $\theta_\phi$, respectively. The two neural networks are denoted as $\psi(x_k|\theta_\psi)$ and $\phi(x_k,u_k,p_k|\theta_\phi)$.

Both neural networks are configured as feedforward neural networks. The forward propagation of $\psi(x_k|\theta_\psi)$ is expressed as
\begin{equation}
    \psi(x_k|\theta_\psi) = h_{v,\psi}\circ h_{v-1,\psi}\circ \dots \circ h_{1,\psi}(x_k)
\end{equation}
where $v$ is the number of layers; $h_{i,\psi}(h_{i-1,\psi})=\sigma_i(W_{i-1,\psi}h_{i-1,\psi}+b_{i-1,\psi})$, with  $\sigma_i$, $i=1,\ldots, v$, being the activation functions and $\theta_{\psi} = (W_{1,\psi},\dots, W_{v,\psi},b_{1,\psi},\dots,b_{v,\psi})$ being the trainable parameters of the neural network. $\phi(x_k|\theta_\phi)$ has a similar structure as $\psi(x_k,u_k,p_k|\theta_\psi)$, and its expression is omitted.
Rectified linear units (ReLU) are utilized as the activation functions for both $\psi(x_k|\theta_\psi)$ and $\phi(x_k,u_k,p_k|\theta_\phi)$. 
The number of layers and neurons in the neural networks can be determined through trial and error, considering the complexity and scale of the underlying system.

\subsection{Koopman model training}
The parameters of the Koompan model are optimized using batch data $\mathcal{D}=\left\{\left(x_k, u_k,p_k\right)_{k=1,\ldots, n}\right\} $ containing $n$ samples of the original states and the known inputs.
The trainable parameters of the proposed modeling approach include the Koopman model matrices $A$, $B$, and $C$, and the neural network parameters $\theta_\psi$ and $\theta_\phi$. These trainable parameters are optimized to minimize the weighted multi-step-ahead prediction errors for the states in the original state space. Specifically, the optimization problem for model training is formulated as follows:
\begin{subequations}\label{eq:loss}
\begin{align}
\min_{A,B,C,\theta_\psi,\theta_\phi}\text{    }& \mathbb{E}_{\mathcal{D}_j}\sum_{j=k}^{k+H} \Vert \hat x_{j|k}-x_{j} \Vert_2^2 \label{opti_loss}\\
\text{s.t. } \hat z_{j+1|k}&= A\hat z_{j|k}+B\left[u_{j}^{\text{T}},p_{j}^{\text{T}},\phi^{\text{T}} (\hat x_{j|k},u_j,p_j)\right]^{\text{T}}\\
\hat z_{k|k}&=\psi(x_k)\\
\hat x_{j|k}&=C \hat z_{j|k}
\end{align}
\end{subequations}
where $\hat z_{j|k}$ denotes the prediction of the state vector in the lifted space for time instant $j$ at time instant $k$;
$\hat x_{j|k}$ denotes the predicted state vector for time instant $j$ at time instant $k$;
% ; (with a slight abuse of notation).
$H$ is length of the multi-step-ahead model prediction horizon; 
% the prediction horizon. 
$\mathcal D_j=\left\{\left(x_j, u_j,p_j\right)_{j=k,\ldots, k+H}\right\}$, which is a subset of $\mathcal D$, contains the samples from time instant $k$ to time instant $k+H$.
{\color{black}
\begin{rmk}
EDMD\cite{williams2015data}, which exploits manually selected observable functions to account for the nonlinear mapping between the original and the lifted spaces, is a practical approach for many nonlinear processes. Meanwhile, for large-scale nonlinear processes, manually selecting a large number of appropriate observable functions can be very challenging. In these case scenarios, it can be more appropriate to use neural networks to account for the observable functions and train the neural networks using process data. In this work, the challenging and time-consuming task of manual selection of the observable functions is bypassed with the introduction of neural networks.
\end{rmk}}

\section{Koopman-based MPC}
In this section, we present a Koopman MPC control formulation based on the established Koopman model with input augmentation for the nonlinear process in \eqref{paper:1:nonlinear:p}. {\color{black} Additionally, we propose an iterative implementation method for the formulated Koopman MPC, using which we only need to solve quadratic programming despite the nonlinearity of the established Koopman model in the form of \eqref{eq3}.}
% that aims to solve the online optimization problem iteratively using convex optimization, particularly addressing the nonlinearity of the established Koopman model in the control input.
% First, the MPC problem based on the learned Koopman model is formulated; then the optimization procedure of the MPC problem is described.

\subsection{Formulation of the optimization problem}
%change the name of MPC

Based on the established Koopman model with input augmentation in the form of \eqref{eq3}, a Koopman-based MPC controller is developed for the nonlinear process in \eqref{paper:1:nonlinear:p}. Specifically, the optimization problem associated with this controller is formulated as follows:
\begin{subequations}    \label{paper1:iterMPC_all}
    \begin{align}
    \min_{u^*_{k|k},\dots,u^*_{k+N-1|k}}\quad  &\sum_{j=k}^{k+N-1}\Vert  C\hat {z}_{k|k}-x_s\Vert_Q^2+\Vert  {u}_{j|k}-u_s\Vert_R^2\label{paper1:iterMPC_all:cost}\\
    \mathrm{s.t.}\quad &\hat{z}_{{j+1}|k}= A\hat{z}_{{j}|k}+B\left[u_{j|k}^{\text{T}},p_j^{\text{T}},\phi^{\text{T}} (C\hat {z}_{k|k},u_{j|k},p_j)\right]^{\text{T}}\label{paper1:iterMPC_all:model}\\
    % \label{eq:state-space_C}  
    % % &u_{j|k}=\left[u_{{c,j}|k}^{\text{T}},u_{{p,j}|k}^{\text{T}}\right]^{\text{T}}\\
    % &\hat{x}_{j|k} = C\hat{z}_{j|k}\\
    &\hat {z}_{k|k} = \psi(x_k)\label{paper1:iterMPC_all:initialstate}\\
    &C\hat{z}_{j|k}\in \mathcal{X}\label{paper1:iterMPC_all:statecons}\\
    &u_{j|k}\in \mathcal{U}, j=k,\dots,k+N-1\label{paper1:iterMPC_all:inputcons}.
    \end{align}
\end{subequations}% where $\|\cdot\|_P$ is the 2-norm of states and control input weighted by the given matrix $P$. $Q$, $R$, $P$ are positive definite weighting matrices, $x_s$ is the tracking target and $u_s$ is the input of stable state.
% The first element of the optimal solution $\left[u^*_{c,k},\dots,u^*_{c,k+H-1}\right]$, 
% $u^*_{c,k}$, is used as the control input at time step $k$.
where $u^*_{k|k},\dots,u^*_{k+N-1|k}$ is the sequence of optimal control inputs that can be obtained via solving \eqref{paper1:iterMPC_all} at time instant $k$; $N$ is the length of the control horizon; $\left\| \cdot \right\|^2_Q$ denotes the square of the weighted Euclidean norm of a vector, computed as $\left\| x \right\|^2_Q= x^{\text{T}}Qx$; $Q$ and $R$ are positive-definite weighting matrices. 
In \eqref{paper1:iterMPC_all}, \eqref{paper1:iterMPC_all:model} serves as the model constraint; (\ref{paper1:iterMPC_all:initialstate}) initializes the state prediction sequence in the lifted state space; (\ref{paper1:iterMPC_all:statecons}) and (\ref{paper1:iterMPC_all:inputcons}) impose constraints on the states and control inputs, respectively. At each sampling instant $k\geq 0$, as the optimal control input sequence is obtained, the first element $u^*_{k|k}$ is applied to the nonlinear process in \eqref{paper:1:nonlinear:p} as the control action for closed-loop process operation. 

It is noted that $\phi$ is a nonlinear function of the known inputs. Therefore, the formulation in \eqref{paper1:iterMPC_all} presents a non-convex optimization problem. To leverage the benefits of Koopman-based MPC in transforming a nonlinear optimal control problem into convex optimization \cite{korda2018linear}, we propose an iterative implementation solution for the formulated non-convex Koopman-based MPC in \eqref{paper1:iterMPC_all}, as detailed in the following subsection.

\subsection{Iterative convex Koopman MPC}

To circumvent the need to solve the non-convex optimization problem in (\ref{paper1:iterMPC_all}), we propose an iterative implementation method. This method allows for leveraging the benefits of Koopman-based MPC methods, e.g., \cite{han2023robust,zhang2023reduced,klus2020data,calderon2021koopman},
in tackling nonlinear control problems via convex optimization. Specifically, instead of employing nonlinear optimization solvers to solve (\ref{paper1:iterMPC_all}) directly, we approximate the optimal solution $u^*_{k|k},\dots,u^*_{k+N-1|k}$ of the optimization problem in (\ref{paper1:iterMPC_all}) by formulating a convex optimization problem and solve it iteratively within each sampling period.

Matrix $B$ in (\ref{eq3}) can be divided into three blocks, that is,
$B =  \left[
    \begin{array}{c: c: c}
        B_u&B_p&B_{\phi}\\
    \end{array}
\right]$. Accordingly, the Koopman model in (\ref{eq3}) can be expressed as:
\begin{equation}\label{paper1:Koopman:expanded}
\begin{aligned}
\hat{z}_{k+1}&= A\hat{z}_{k}+B_u u_{k}+B_p p_{k}+B_{\phi} \phi (x_k,u_k,p_k)\\
\hat{x}_{k}&= C{\hat z}_{k}\\
\end{aligned}
\end{equation}

Based on (\ref{paper1:Koopman:expanded}) and the non-convex control formulation in (\ref{paper1:iterMPC_all}), we formulate an iterative convex optimization problem as follows: 

\begin{subequations}\label{paper1:iterMPC}
    \begin{align}    
    \min_{u^{[l]}_{k|k},\dots,u^{[l]}_{k+N-1|k}}\quad  &\sum_{j=k}^{k+N-1}\Vert  C\hat{z}^{[l]}_{j|k}-x_s\Vert_Q^2+\Vert  {u}^{[l]}_{j|k}-u_s\Vert_R^2\\
    \mathrm{s.t.}\quad &\hat{z}^{[l]}_{{j+1}|k}= A\hat{z}^{[l]}_{{j}|k}+B_u u^{[l]}_{j|k}+B_p p_{j}+B_{\phi} \phi \Big(C {\hat z}^{[l-1]}_{j|k},u^{[l-1]}_{j|k},p_{j}\Big)\label{paper1:iterMPC:model}  \\
    &\hat {z}^{[l]}_{k|k} = \psi(x_k)\label{paper1:iterMPC:initialstate}\\
    & C\hat{z}^{[l]}_{j|k}\in \mathcal{X}\label{paper1:iterMPC:statecons}\\
    &u^{[l]}_{j|k}\in \mathcal{U},~~~ j=k,\dots,k+N-1.\label{paper1:iterMPC:inputcons}
    \end{align}
\end{subequations}
(\ref{paper1:iterMPC}) is required to be solved iteratively for $l_{\max}$ iteration steps within each sampling period. In (\ref{paper1:iterMPC}), the superscript $l$, $l=1,\ldots,l_{\max}$, denotes the current iteration step; $u^{[l]}_{k|k},\dots,u^{[l]}_{k+N-1|k}$ represents the sequence of control inputs generated in the $l$th iteration step at time instant $k$; $\hat{x}^{l}_{j|k}$ is the prediction of the state for time instant $j$, obtained in the $l$th iteration step at sampling instant $k$. 

In (\ref{paper1:iterMPC:model}), ${\hat z}^{[l-1]}_{j|k}$ and ${u}^{[l-1]}_{j|k}$ are known fixed parameters obtained from solving the most recent optimization problem associated with (\ref{paper1:iterMPC}). 
For $l>1$, ${u}^{[l-1]}_{j|k}$, $j=k,\ldots,k+N-1$, is the optimal control input generated in the previous iteration step (i.e., $(l-1)$th iteration step) at time instant $k$. When $l=1$, ${u}^{[0]}_{j|k}$ is determined based on the optimal control input sequence generated in the final iteration step at time instant $k-1$, as follows:
\begin{equation}\label{paper1:input:previous}
{u}^{[0]}_{j|k} =\left\{ \begin{array}{l}\hspace{-1.6mm}
 {u}^{[l_{\max}]}_{j|k-1},~~~~~~~~~~~{\text{for}}~ k\leq j \leq k+N-2~~\\[0.2em]
\hspace{-1.6mm}{u}^{[l_{\max}]}_{k+N-2|k-1},~~~~{\text{for}}~ j=k+N-1~~
\end{array} \right.
\end{equation}
The sequence of predicted states ${\hat z}^{[l-1]}_{j|k}$, $j=k,\ldots,k+N-1,$ is generated based on the model in (\ref{paper1:iterMPC:model}) using the input sequence containing ${u}^{[l-1]}_{i|k}$, $i=k,\ldots,k+N-2$. 

The optimal solution obtained in the final iteration step, denoted by $u^{[l_{\max}]}_{k|k},\dots,u^{[l_{\max}]}_{k+N-1|k}$, is treated as the optimal control input sequence for the current time instant $k$. The first element of this control sequence, denoted by $u^{[l_{\max}]}_{k|k}$, is applied to the nonlinear system 
in (\ref{paper:1:nonlinear:p}). 
An implementation algorithm that details the iterative execution of the proposed approach is provided in Algorithm~\ref{paper1:alg:2} below.

Despite the nonlinearity of function $\phi$ in (\ref{paper1:iterMPC:model}),  $C {\hat z}^{[l-1]}_{j|k}$, $u^{[l-1]}_{j|k}$, and $p_{j}$ are all fixed known parameters. Only $u^{[l]}_{j|k}$, $j=k,\ldots, k+N-1$, are the decision variables of the optimization problem in (\ref{paper1:iterMPC}). Therefore, (\ref{paper1:iterMPC}) represents a convex optimization problem, which can be solved efficiently using quadratic programming solvers. 
\begin{algorithm}[t]
\caption{Iterative implementation of the proposed input-augmented Koopman MPC}\vspace{3mm}
\label{paper1:alg:2}
At time instant $k\geq 1$, the controller conducts the following steps:
\begin{enumerate}
    \item[1.] Generate ${u}^{[0]}_{j|k}$, $j=k,\ldots,k+N-1$  following \eqref{paper1:input:previous}, obtain ${\hat z}^{[0]}_{j|k}$, $j=k,\ldots,k+N-1$, and the known disturbances $p_j$, $j=k,\ldots,k+N-1$.
    \item[2.] Set $l$ = 1.
    \begin{enumerate}
        \item [2.1.] Solve the optimization problem in \eqref{paper1:iterMPC} to generate ${u}^{[l]}_{j|k}$, $j=k,\ldots,k+N-1$, and ${\hat z}^{[l]}_{j|k}$, $j=k,\ldots,k+N-1$.
        \item[2.2.]
        {\emph{\textbf{If}}} $l=l_{\mathrm{max}}$, do:
    \begin{itemize}
        \item ${u}^{[l_{\max}]}_{k|k}$ is applied to the nonlinear system in \eqref{paper:1:nonlinear:p}.
    \end{itemize}
     {\emph{\textbf{Else}}}, do:
     \begin{itemize}
         \item Set $l=l+1$. Go to step 2.1.
     \end{itemize}
    \end{enumerate}
    %\item[5.] The DMHE-2 estimator receives a sequence of state estimate $\hat{x}_{k}^{i}$, $i\in\mathbb{N}$ from the estimtors.
     \item[3.]\label{final} Set $k=k+1$. Go to step 1.
     \end{enumerate}
\end{algorithm}

\begin{rmk}
The total number of iteration steps for each sampling instant, denoted as $l_{\max}$, is user-specified and can be determined through trial-and-error. An alternative approach involves incorporating a stopping criterion. This criterion compares the optimal control inputs generated in two consecutive iteration steps. If the difference between the two optimal control sequences, evaluated based on metrics such as the Euclidean norm, is sufficiently small, then the iterative execution can be terminated for the current sampling instant, and the optimal control is generated. This adaptive approach can allow for variations in the number of iteration steps as time progresses. As the process operation approaches steady-state levels, the number of iteration steps may decrease, which will further enhance online computational efficiency.
\end{rmk}

\section{Application to a reactor-separator process}
\label{experiment1}
\begin{figure}[ht]
  \centering
  \includegraphics[width=1\textwidth]{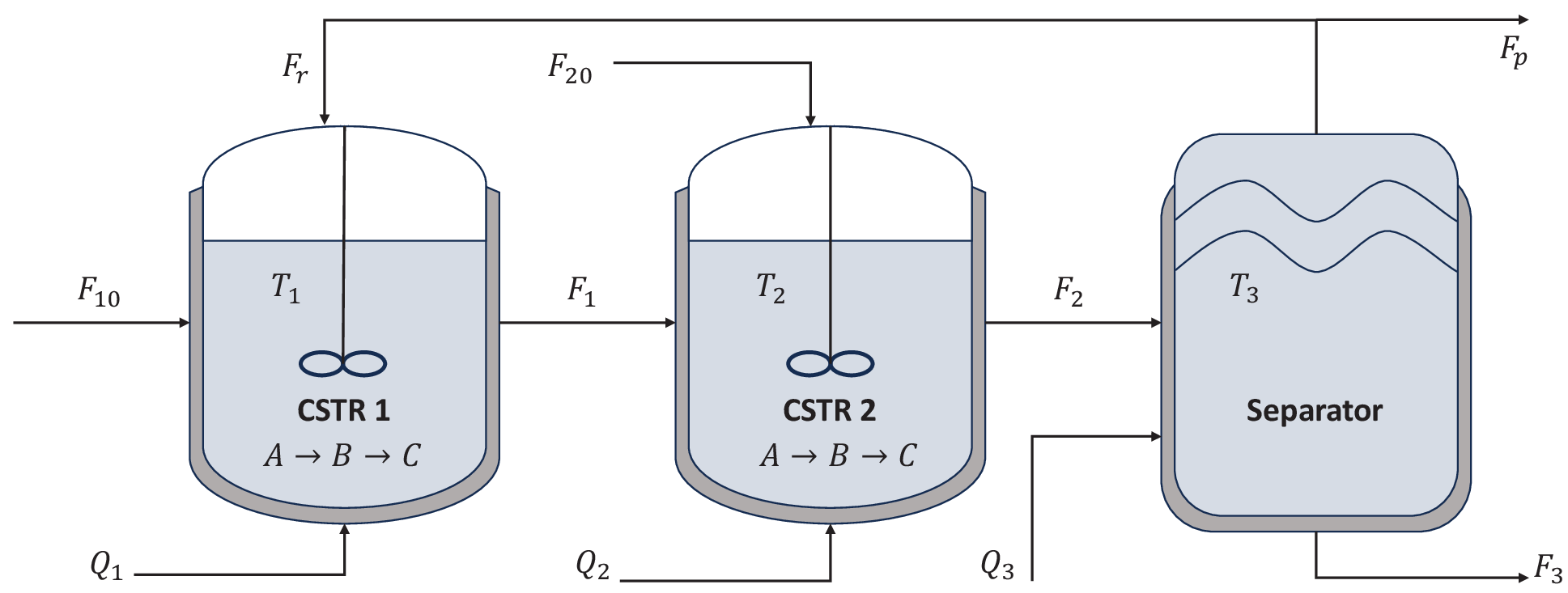}
  \caption{A illustrative diagram of the reactor-separator process. } 
  \label{figure:CSTR}
\end{figure}
\subsection{Process description}
In this section, the proposed method is applied to a benchmark chemical process via simulations.
This reactor-separator process comprises two continuous stirred tank reactors (CSTRs) and a flash tank separator. A schematic of this process is presented in Figure~\ref{figure:CSTR}. 
The process involves two chemical reactions. The first reaction generates the desired product $B$ by converting the reactant $A$. At the same time, in the second reaction, a portion of $B$ is converted into the side product $C$. A feed flow that takes pure reactant $A$ enters the first reactor (i.e., CSTR 1) at a flow rate of $F_{10}$. The outflow of CSTR 1 at flow rate $F_1$ and an additional fresh feed of pure $A$ at flow rate $F_{20}$ are the inlets to CSTR 2. The effluent of CSTR 2 flows into the separator at flow rate $F_2$. In the separator, a recycle stream at a flow rate of $F_r$ is directed to the first reactor for further reaction. Each of the three vessels is equipped with a jacket, which provides/removes heat at a heating input rate $Q_i, i=1,2,3$. A more detailed description of this process can be found in \cite{liu2008two,yin2019subsystem}.

The state variables include the mass fractions of substances $A$ and $B$ in the three vessels, denoted by $x_{Ai}$ and $x_{Bi}, i = 1,2,3$, and the temperatures in the three vessels, denoted by $T_i, i = 1,2,3$.
The state vector of this process is $x = [x_{A1},x_{B1},T_1,x_{A2},x_{B2},T_2,x_{A3},x_{B3},T_3]^{\text{T}}$.
The control input vector is $u =\big[Q_1, Q_2, Q_3\big]^{\text{T}}$. The control inputs are subject to constraints: $ \left[0,0,0\right]^{\text{T}} \leq u \leq \left[4.87,1.68,4.87\right]^{\text{T}}\times 10^6\,\text{kJ}/\text{h}$.
Based on the mass and energy balances, a nonlinear model that consists of nine ordinary differential equations (ODEs) has been established to characterize the dynamic behaviors of the variable. This dynamic model and the values of the model parameters can be found in \cite{zhang2013distributed}.
No known disturbances $p$ are involved in this process. 

\subsection{Modeling results}

A dataset $\mathcal{D}$ is generated by the nonlinear model based on the ODEs. 
The states are measured with a sampling period of $\Delta=0.005\,\text{h}$. 
A sampling trajectory containing 12000 samples is generated with the initial state $x_{init}=\left[0.1155, 0.6235,497.3\,\text{K}, 0.1367, 0.6053, 489.8\,\text{K}, 0.0396, 0.5504, 491.8\,\text{K}\right]^{\text{T}}$.
At sampling instant $k$, the control inputs signal is generated following $u_k=\bar{u}+\epsilon_u$, where $\bar{u}$ is randomly generated following uniform distribution subject to constraints on the inputs and remains constant for the subsequent 20 sampling periods. The noise $\epsilon_u$ is generated following a normal distribution $\epsilon_u\sim \mathcal{N}(\mathbf{0},\sigma_u)$ and is updated at each time instant, with $\sigma_u=\left[3.25,1.12,3.25\right]\times 10^4$. Process disturbances are generated following a multivariate normal distribution $\mathcal{N}(\mathbf{0},\sigma_\epsilon^2)$ with $\sigma_\epsilon=\left[0.01, 0.01, 0.50, 0.01, 0.01, 0.50, 0.01, 0.01, 0.50\right]$. The process disturbances for the states are bounded within the $[-5,5]$ range and then added to the respective state variables.
The entire dataset is divided into three segments: 9000 samples for training, 1000 samples for validation, and 2000 samples for tests.

In the training process, normalization is conducted such that the data for each state and input variable have a mean of 0 and a standard deviation of 1. 
The model prediction horizon is chosen as 40. 
A Koopman model is established using the proposed DKOIA modeling approach. Additionally, the proposed DKOIA method is compared with a baseline method -- Deep Koopman Operator (DKO) \cite{lusch2018deep,han2020deep}, which does not involve input augmentation.
To ensure fair comparisons, the structures of the neural networks accounting for the observable functions are made identical.
The hyperparameters used for training are specified in Table \ref{table:CSTR parameters}.

\begin{table}[t]                                                                                   
	\centering
	\caption{Hyperparameters of the training process for the reactor-separator process}
	\label{table:CSTR parameters}
        \renewcommand\arraystretch{1.2}
	\setlength{\tabcolsep}{5mm}{
		\begin{tabular}{cc}
			\toprule
			\textbf{Hyperparameters}&\textbf{Value}\\
			\midrule 
            % Prediction horizon $H$ & 20\\
            Training epochs&400\\
			Batch  size& 128\\
                Model prediction horizon & 40\\
                Learning rate& $10^{-3}$\\
                Activation function& ReLU\\
                $l_2$ norm regularization coefficient& 0.1\\
                Input dimension of $\psi$ & 9\\
                Output dimension of $\psi$ & 13\\
                Input dimension of $\phi$ & 12\\
                Output dimension of $\phi$ & 6\\
                Structure of $\psi$ & (32,64,32)\\
                Structure of $\phi$ & (16,32,16)\\
			\bottomrule
	\end{tabular}}
\end{table}

After training for 400 epochs, both algorithms converge on the training, validation, and test sets. For the 40-step-ahead prediction task, the average prediction errors of the DKOIA model and the DKO model on the validation set are $2.13\times 10^{-4}$ and $8.14\times 10^{-3}$ for each step, respectively. On the test dataset, the average prediction errors are $3.08\times 10^{-4}$ and $9.96\times 10^{-3}$ for each step. {\color{black}While the states of this chemical process do not have nonlinear dependence on the known inputs based on the ODEs characterizing the process dynamics, the proposed method is able to provide significantly reduced model prediction errors
as compared to the DKO-based modeling approach.}

\begin{figure}[t]
  \centering
  \includegraphics[width=0.95\textwidth]{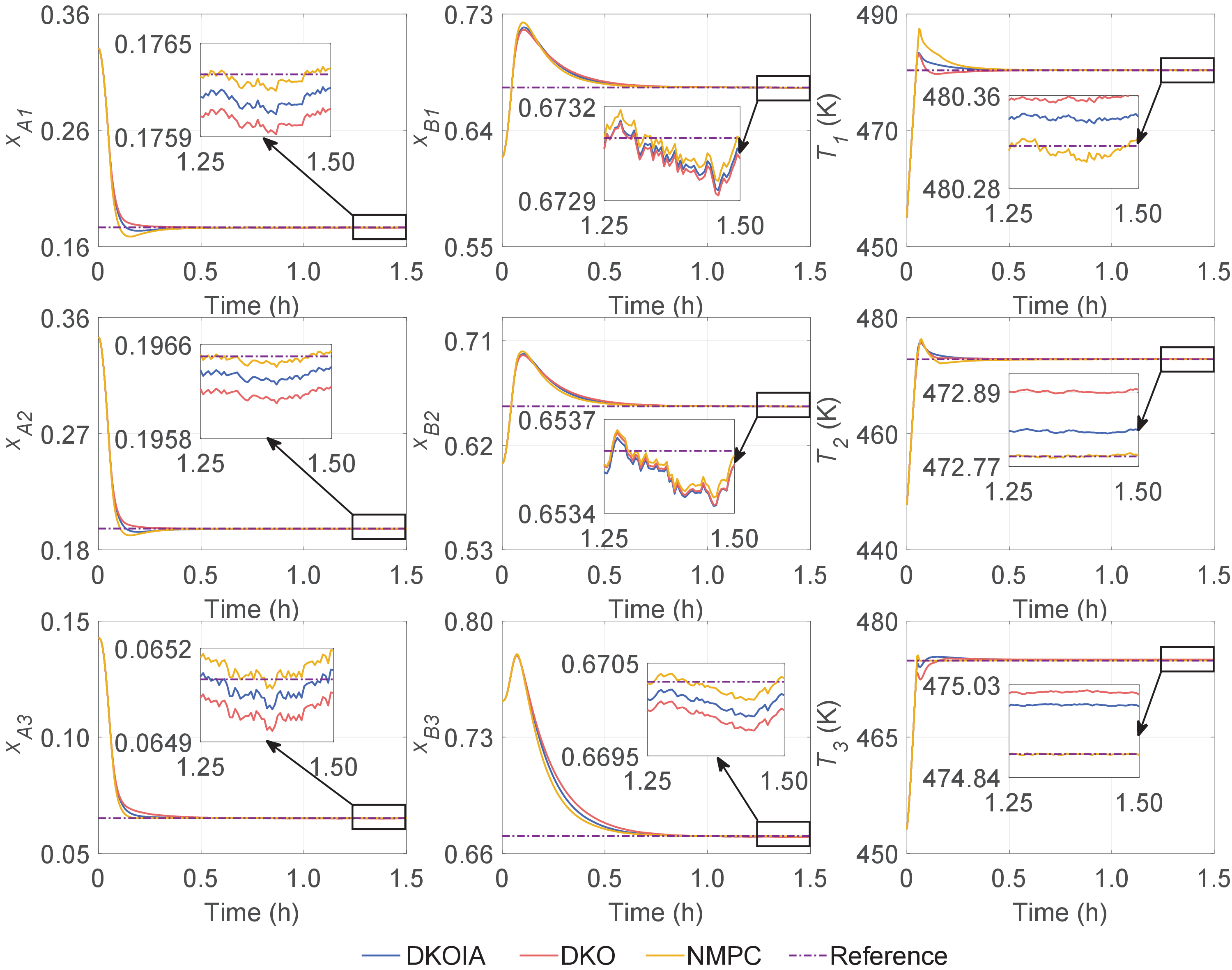}
  \caption{Closed-loop state trajectories for the reactor-separator process under the DKOIA-based controller, the DKO-based controller, and NMPC.} 
  \label{figure:result_all}
\end{figure}

\begin{figure}[t]
  \centering
  \includegraphics[width=0.95\textwidth]{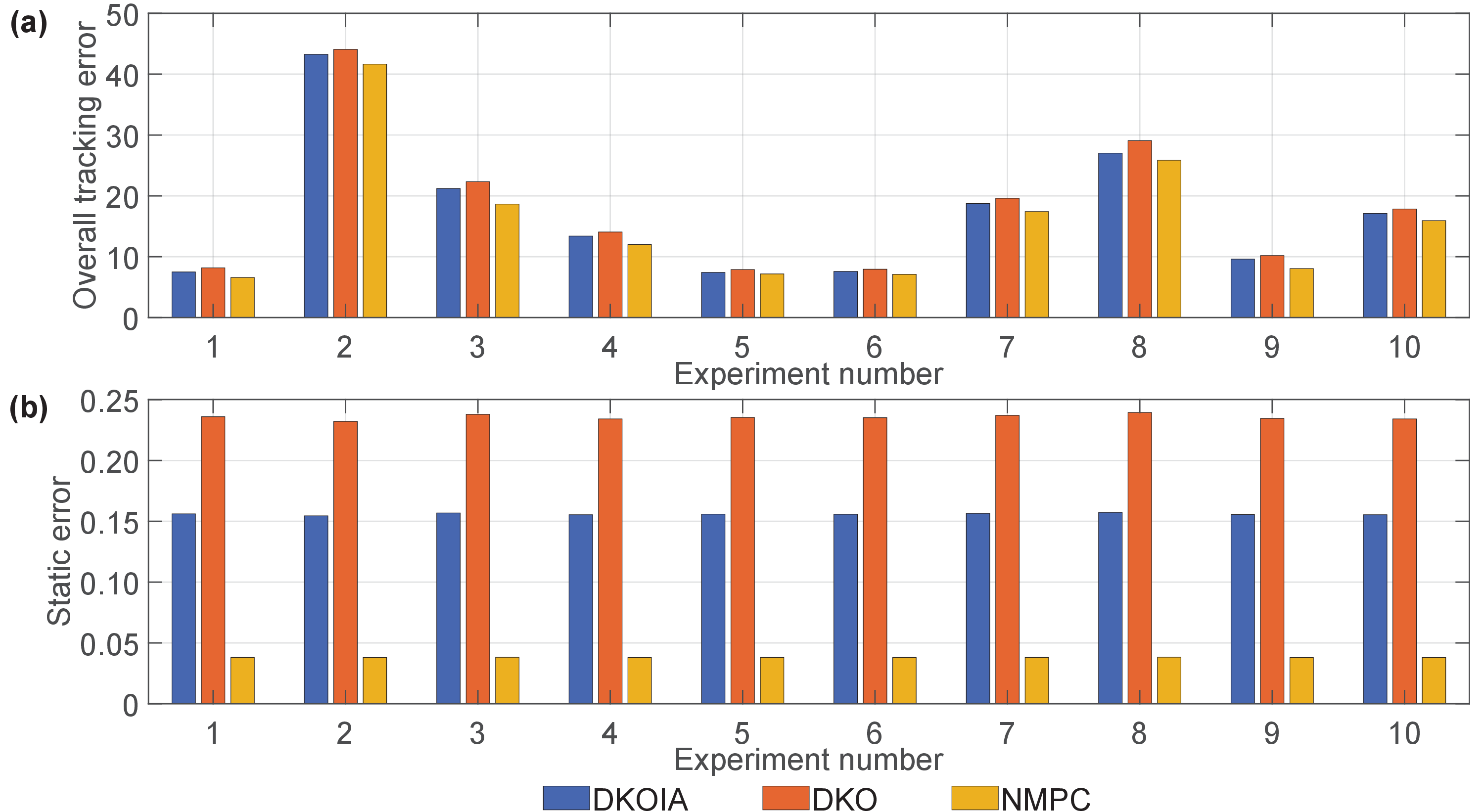}
  \caption{Comparisons of the control performance under the DKOIA-based controller, the DKO-based controller, and NMPC: (a) overall tracking errors and (b) static errors.} 
  \label{figure:result_bar}
\end{figure}

\subsection{Control performance}
In this section, we evaluate the control performance for this chemical process by using three methods, including the proposed DKOIA-based method, the DKO-based method in \cite{lusch2018deep,han2020deep}, and nonlinear MPC (NMPC) based on the first-principles nonlinear model of the considered chemical process.
The control objective is to drive the process operation towards and maintain it at steady-state set-point  $x_s=\left[0.18, 0.67, 480.32\,\text{K}, 0.20, 0.65, 472.79\,\text{K}, 0.07, 0.67, 474.89\,\text{K}\right]^{\text{T}}$.
The corresponding steady-state input is $u_s =\left[2.90,1.00,2.90\right]\times 10^6\,\text{kJ}/\text{h}$.
% The weighting matrices are made the same for the three controllers, which are $Q=diag(0.5, 0.5, 1.5, 0.5, 0.5, 1.5, 0.5, 0.5, 1.5)$, and $R=diag(0.05,0.05,0.05)$. 
The weighting matrices of the three controllers are carefully tuned for improved control performance. For the DKOIA-based controller, the weighting matrices are $Q_{DKOIA}=\text{diag}(1.5, 0.1, 3.3, 2.4 ,0.4, 1.5, 1.5, 0.1, 3.3)$, and $R_{DKOIA}=\text{diag}(0.002,0.002,0.0001)$. For the DKO-based controller, the weighting matrices are $Q_{DKO}=\text{diag}(1.7, 0.2, 1.3, 1.7 ,0.2, 0.5, 1.9, 0.2, 2.1)$, and $R_{DKO}=\text{diag}(0.01,0.005,0.001)$. For NMPC, the weighting matrices are $Q_{NMPC}=\text{diag}(0.5, 0.5, 1.5, 0.5 ,0.5, 1.5, 0.5, 0.5, 1.5)$, and $R_{NMPC}=\text{diag}(0.005,0.005,0.005)$. 
The control horizon is 40 for each controller. $l_{max}$ is chosen as 2 for the DKOIA-based controller. 
% 10 experiments with different initial states are conducted, and the simulation duration is 1.5 h for each experiment. 
% the prediction horizon $H$ is 20 for each controller, and the control horizon is 10, which means the control input will not change after 10 steps.

% The trajectories of the nine states based on three control methods are shown in Figure~\ref{figure:result_all}. 
% The DKOIA-based controller drives the process operation towards the steady state faster than the controller based on DKO.
With initial state $x_{init}=\left[0.33, 0.62, 454.98\,\text{K}, 0.34, 0.605, 447.76\,\text{K}, 0.14, 0.75, 453.09\,\text{K}\right]^{\text{T}}$,
closed-loop trajectories of the nine state variables within a 1.5-hour operation period given by the three controllers are shown in Figure~\ref{figure:result_all}. The three controllers can accomplish the set-point tracking task. The proposed DKOIA-based controller has smaller tracking errors after the convergence of the system compared with the DKO-based controller. 
% The control performance is evaluated by the root mean squared errors (RMSE) between the state and the steady-state set-point $xs$.
At each time instant, the root mean squared errors (RMSEs) between the state $x_k$ and the steady-state set-point $x_s$ are calculated to evaluate the control performance.
Since the nine states have different magnitudes, the RMSEs are calculated based on the states after normalization. 
The overall tracking errors are the sum of the RMSEs within the whole simulation duration, while the static tracking errors are the mean of the RMSE within the last 0.25-hour simulation time.
Ten experiments with different initial states are conducted, and the overall tracking errors and static tracking errors of each experiment are shown in Figure~\ref{figure:result_bar}. 
Among the ten experiments, the mean values of overall tracking errors for the controllers based on DKOIA, DKO, and NMPC are 17.29, 18.12, and 16.05, respectively. 
The DKOIA-based controller reduces the overall tracking error by 4.53$\%$ for this process, as compared with the DKO-based method. % From the result of static error, the nonlinear solver with a first-principles model has the least static error of 0.0141, and  DKO has the largest static error of 0.0504. The DKOIA reduces $49.31\%$ static error compared with DKO, which has a static error of 0.0255. 
The mean values of the static tracking errors for the controllers based on DKOIA, DKO, and NMPC are $1.56\times 10^{-1}$, $2.36\times 10^{-1}$, and $3.81\times 10^{-2}$, respectively. The proposed method reduces the static errors by 33.79$\%$ as compared to the DKO-based controller.
{\color{black} 
NMPC provides better control performance than the controllers based on DKOIA and DKO
in terms of both overall tracking errors and static errors, since the NMPC assumes that a perfect first-principles nonlinear dynamic process model is available. This assumption, however, can be very challenging to satisfy in real-world applications.}

\section{Application to a wastewater treatment plant}
\label{experiment2}

\subsection{Process description}
We further consider a biological water treatment process based on the Benchmark Simulation Model No. 1 (BSM1)\cite{alex2008benchmark} to evaluate the performance of the proposed method on large-scale complex systems.
\begin{figure}[t]
  \centering
  \includegraphics[width=\textwidth]{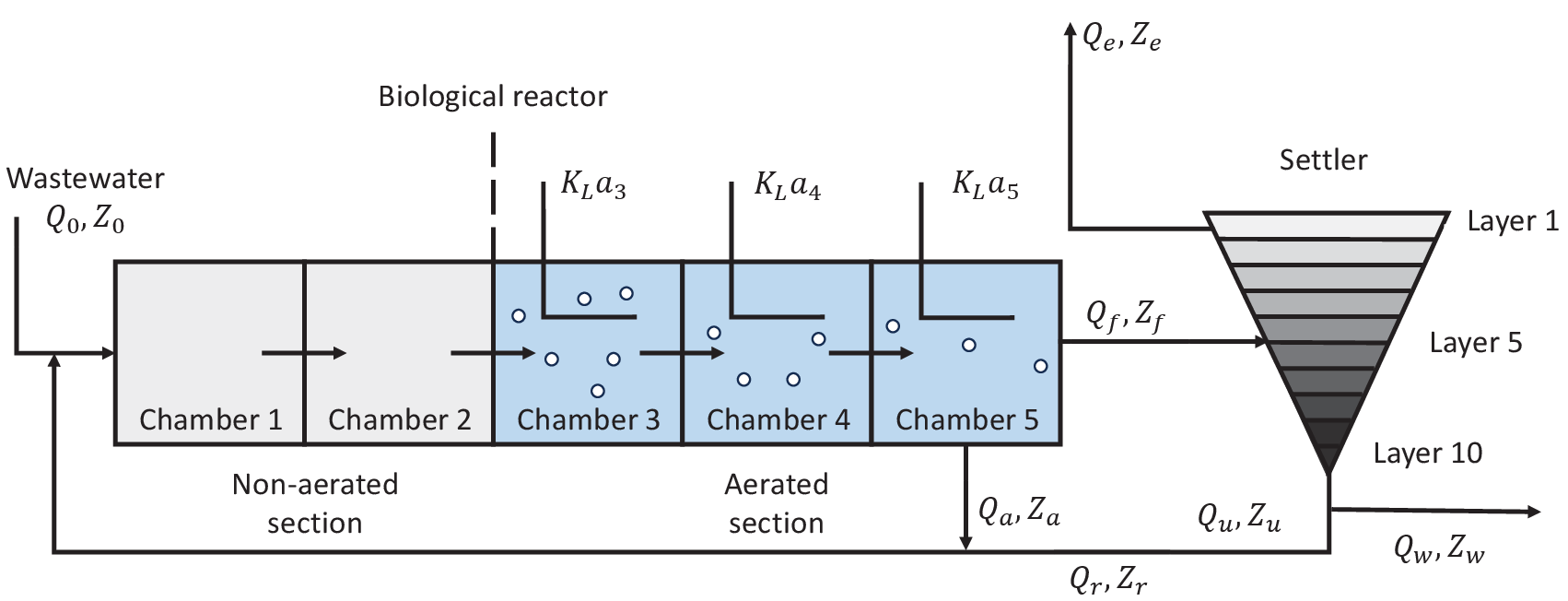}
  \caption{A schematic of the wastewater treatment plant. } 
  \label{figure:wastewaster}
\end{figure}
A schematic of this wastewater treatment process (WWTP) is given in Figure ~\ref{figure:wastewaster}. As shown in Figure~\ref{figure:wastewaster}, the process comprises a biological reactor and a secondary settler. The biological reactor has five chambers: the first two are anoxic and form the non-aerated section, and the remaining three are aerobic and form the aerated section.
In the non-aerated section, nitrates are converted into nitrogen through denitrification. In the aerated section, nitrification takes place to oxidize ammonia into nitrate.
The settler has ten non-reactive layers, with the fifth being the feed layer. 

Wastewater enters the first chamber of the reactor at flow rate $Q_0$ and concentration $Z_0$.
A portion of the outlet stream from Chamber 5 enters the settler through Layer 5 at flow rate $Q_f$ and concentration $Z_f$. The remainder of the outlet stream from Chamber 5 enters Chamber 1 at flow rate $Q_a$ and concentration $Z_a$. In the settler, purified water leaves the plant from the top layer (i.e., Layer 1 in Figure~\ref{figure:wastewaster}) at flow rate $Q_e$ and concentration $Z_e$. The underflow of the settler, discharged from the bottom layer (i.e., Layer 10), contains two portions: 1) a recycle flow back to the first Chamber 1 of the biological reactor at flow rate $Q_r$ and concentration $Z_r$; 2) the remainder with sedimented sludge is discharged at flow rate $Q_w$ and concentration $Z_w$\cite{alex2008benchmark}.

In the biological reactor, we consider 8 biological reactors that involve 13 compounds. The concentrations of the compounds in the five chambers are the state variables for the reactor, which are defined in Table~\ref{table:table2_wwtp}. The dynamic behaviors of the 13 state variables for each chamber of the reactor are governed by 13 ordinary differential equations (ODEs).
In the settler, $S_O$, $S_{ALK}$, $S_{NH}$, $S_{NO}$, $S_S$, $S_I$, $S_{ND}$ and $X$ are included to represent the states of each layer, where $X$ describes the concentration of the suspended solids, calculated as the sum of $X_S$, $X_I$, $X_{B_H}$, $X_{B_A}$, $X_P$ and $X_{ND}$. The dynamics of the states in each layer are described by 8 ODEs.
In total, 145 ODEs are formulated to describe the dynamic behaviors of the WWTP. 
The control inputs to the WWTP are the internal recycle flow rate $Q_a$ and the oxygen transfer coefficient in the fifth chamber $K_La_5$, which formulates the control input vector as $u =\big[Q_a, K_La_5\big]^{\text{T}}$. The inputs are subject to hard constraints: $0\leq Q_a \leq 92230 {\rm\, m^3\,day^{-1}}$ and $0\leq K_La_5 \leq 240 {\rm \, day^{-1}}$.
Known disturbances $p_k\in \mathbb{R}^{14}$ are composed of the inlet flow rate of the wastewater to Chamber 1 and the concentrations of the 13 compounds in the flow.
A detailed description of this water treatment process, together with the expressions of the nonlinear dynamic model can be found in \cite{alex2008benchmark,yin2018state,busch2013state,yin2018subsystem}. 
 {\color{black} In this process, the states have nonlinear dependence on the inputs according to the first-principles process model.}

\begin{table}[tbp]
	\centering
	\caption{Defintions of the state variables of the wastewater system  }
 \renewcommand\arraystretch{1.15}
	\label{table:table2_wwtp}
	\setlength{\tabcolsep}{1.5mm}{
		\begin{tabular}{cccc}
			\toprule
			\textbf{State}&\textbf{Physical meaning}&\textbf{Unit}\\
			\midrule 
            $S_I$ & Inert soluble organic matter & ${\rm g\,COD\,m^{-3}}$\\
            $S_S$ & Readily biodegradable and soluble substrate & ${\rm g\,COD\,m^{-3}}$\\
            $X_I$ & Inert particulate organic matter& ${\rm g\,COD\,m^{-3}}$\\
            $X_S$ & Slowly biodegradable and soluble substrate& ${\rm g\,COD\,m^{-3}}$\\
            $X_{B_H}$ & Active heterotrophic biomass & ${\rm g\,COD\,m^{-3}}$\\
            $X_{B_A}$ & Active autotrophic & ${\rm g\,COD\,m^{-3}}$\\
            $X_P$ &  Particulate products from biomass decay& ${\rm g\,COD\,m^{-3}}$\\
            $S_O$ & Dissolved oxygen & ${\rm g\,(-COD)\,m^{-3}}$\\
            $S_{NO}$ & Nitrite nitrogen and nitrate & ${\rm g\, N \,m^{-3}}$\\
            $S_{NH}$ & Free and saline ammonia & ${\rm g\, N \,m^{-3}}$\\
            $S_{ND}$ & Biodegradable and soluble organic nitrogen & ${\rm g\, N \,m^{-3}}$\\
            $X_{ND}$ &  Particulate products from biomass decay& ${\rm g\, N \,m^{-3}}$\\
            $S_{ALK}$ & Alkalinity & ${\rm mol \,m^{-3}}$\\
			\bottomrule
	\end{tabular}}
\end{table}

\subsection{ Modeling results}
We evaluate the proposed method under dry weather conditions. To simulate the process and generate data, we utilize the wastewater profile under the dry weather which includes both flow rate and substance concentrations, sourced from \cite{iwa2010iwa}. 
The initial condition of the 145 state variables is made the same as that in \cite{zeng2015economic}.
% The initial conditions are derived by simulating the plant in a closed-loop system for 150 days with constant inputs. Subsequently, the simulation is conducted under dry weather conditions for an additional 14 days, following the suggested approach outlined in \cite{francisco2011model}. 
Throughout the process operation, the inert soluble organic matter in each chamber of the reactor and each layer of the settler maintains a stable level, despite the fluctuations in the wastewater flow and substance concentrations, and changes in the two control inputs. Therefore, 15 states with respect to the inert soluble organic matter are excluded from Koopman-based modeling and control. We aim to build a Koopman model that describes the time evolution of the remaining 130 state variables. 
Based on the wastewater profile, the process is simulated for 14 days.
The sampling period is set as 15 minutes. 
The control inputs signal at sampling instant $k$ is given as $u_k=\bar{u}+\epsilon_u$ for open-loop simulation, 
where $\bar{u}$ is chosen in the form of sine waves as:
\begin{equation}
    u(t) = A_u\sin{(\omega k + \varphi)}+B_u
\end{equation}
The amplitude $A_u$ and bias $B_u$ are chosen to ensure the sine wave fulfills the range in the input constraints. The angular frequency $\omega$ and phase $\varphi$ are chosen randomly to introduce variability into the input signal. Notably, angular frequency $\omega$  is sampled from the interval $[\frac{2\pi}{96},\frac{2\pi}{288}]$ randomly to maintain the period of the sine wave from one day to three days. $\bar{u}$  remains constant for the next 4 steps. The noise $\epsilon_u$ subjects to a normal distribution $\epsilon_u\sim \mathcal{N}(\mathbf{0},\sigma_u)$ and is updated at each time instant, with $\sigma_u=\left[2700, 72\right]$. The combination of $\bar{u}$ and $\epsilon_u$ helps to generate informative data while limiting large fluctuations in the inputs.

\textbf{\begin{table}[htbp]
	\centering
	\caption{Hyperparameters used in model training for the WWTP}
	\label{table:WWTP parameters}
        \renewcommand\arraystretch{1.15}
	\setlength{\tabcolsep}{1.5mm}{
		\begin{tabular}{ccc}
			\toprule
			\textbf{Hyperparameters}&\textbf{Value}\\
			\midrule 
                % Prediction horizon $H$ & 15\\
                Training epochs&500\\
			Batch size& 256\\
                Model prediction horizon & 10\\
                Learning rate& $10^{-3}$\\
                Activation function& ReLU\\
                $l_2$ norm regularization coefficient& 0.1\\
                Input dimension of $\psi$ & 130\\
                Output dimension of $\psi$ & 150\\
                Input dimension of $\phi$ & 146\\
                Output dimension of $\phi$ & 20\\
                Structure of $\psi$ & (512,1024,512)\\
                Structure of $\phi$ & (512,1024,512)\\
			\bottomrule
	\end{tabular}}
\end{table}}

% The sampling period 
In total, 51110 samples are generated. The dataset is divided into three segments: 40888 samples for training, 10222 samples for validation, and 10760 samples for tests.
% separated into two parts: $4\times 10^4$ samples for the training set and $10^4$ samples for the validation set. Moreover, $10^4$ samples are generated to constitute the test set.
The state and input variables are normalized to have a zero mean and a standard deviation of 1.
Hyperparameters selected for the proposed Koopman modeling approach are listed in Table \ref{table:WWTP parameters}.

 \begin{figure}[t]
  \centering
  \includegraphics[width=0.85\textwidth]{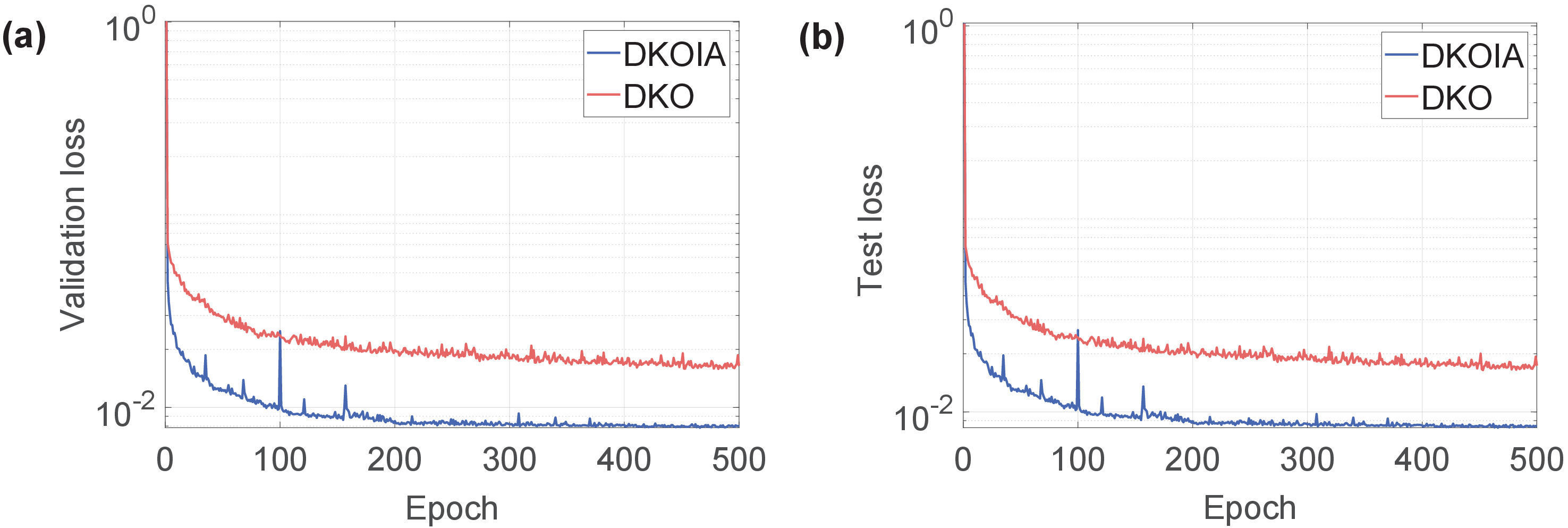}
  \caption{Losses from DKOIA and DKO: (a) the validation set and (b) the test set. }
  \label{figure:loss}
\end{figure}
After training for 500 epochs, both algorithms converge on the training, validation, and test sets. The validation loss and test loss during the training process are shown in Figure~\ref{figure:loss}. 
For the 10-step-ahead prediction task, the average prediction errors of the model from DKOIA and DKO on the validation set are  $8.03\times 10^{-3}$ and $1.65\times 10^{-2}$ for each step, respectively.
The average prediction error of the model from DKOIA on the test set is $8.38\times 10^{-3}$, while the average prediction error of the model from DKO is $1.67\times 10^{-2}$. The proposed method reduces 49.82 $\%$ prediction error on the test set compared with the Koopman modeling without input augmentation. {\color{black} 
The proposed method can effectively handle the nonlinear dependence of the process states on the inputs.}

\begin{figure}[t]
  \centering
  \includegraphics[width=0.9\textwidth]{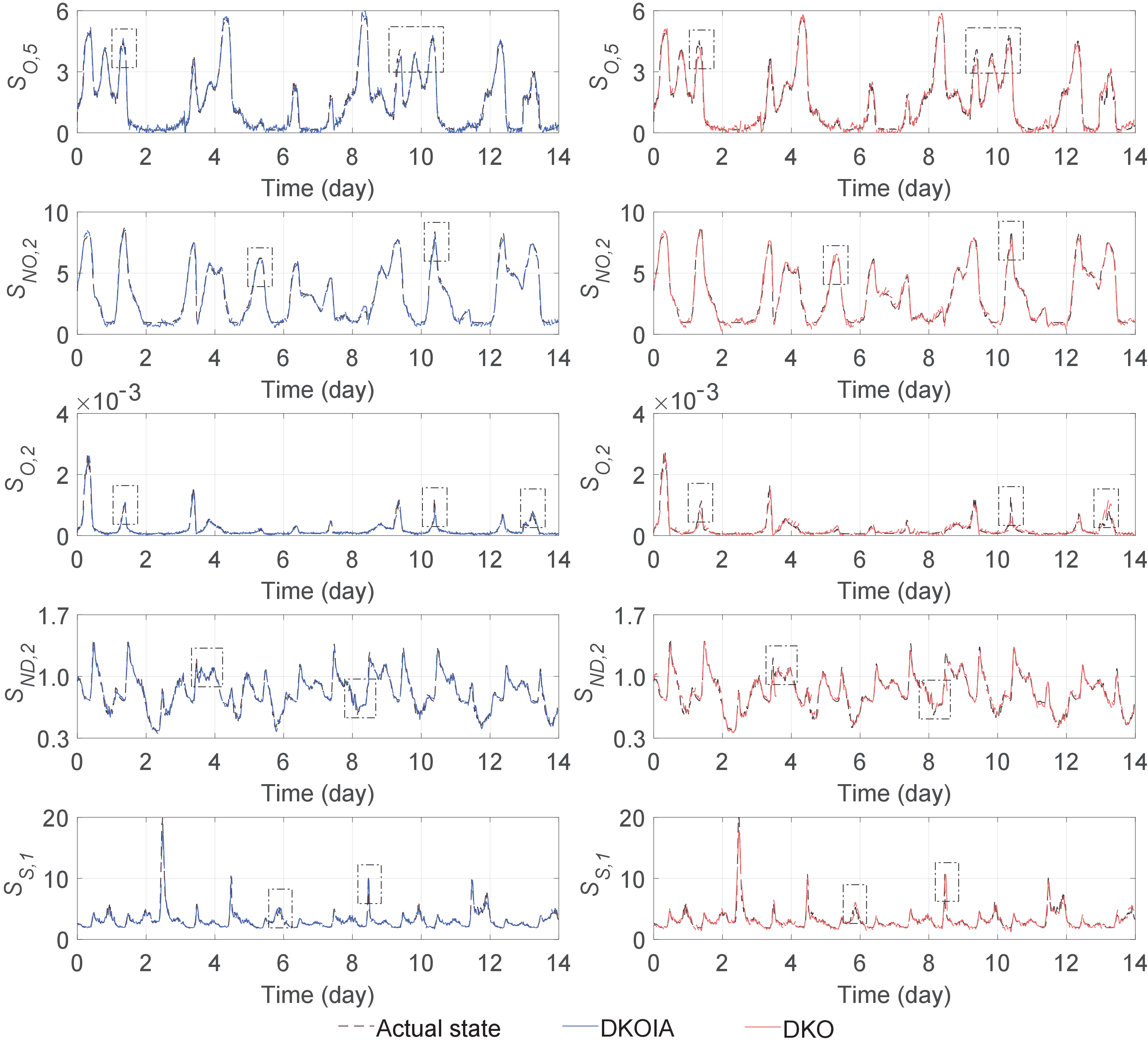}
  \caption{Trajectories of the actual state and the state predictions provided by the DKOIA-based and DKO-based models.
  } 
  \label{figure:compare_pred}
\end{figure}

\begin{table}[htbp]
	\centering
	\caption{Prediction errors of selected state variables}
	\label{table:prediction results}
        \renewcommand\arraystretch{1.15}
	\setlength{\tabcolsep}{1.5mm}{
		\begin{tabular}{cccccc}
			\toprule
            Methods & $S_{O,5} $ &$S_{NO,2} $ & $S_{O,2}$ & $S_{ND,2}$& $S_{S,1}$\\
			\midrule 
            DKOIA & 0.1279 & 0.2283 & $3.9331\times 10^{-5}$ &0.0234&0.3774\\
            DKO & 0.1800 & 0.3216 & $8.9945\times 10^{-5}$ &0.0449&0.7579\\
			\bottomrule
	\end{tabular}}
\end{table}

Splicing actual state and 10-step-ahead predictions results of the same trajectory, part of the predicted states are presented in Figure~\ref{figure:compare_pred} to illustrate the performance of the two methods, which are the dissolved oxygen concentration in Chamber 5 (denoted by $S_{O,5}$), the nitrite nitrogen and nitrate in Chamber 2 (denoted by $S_{NO,2}$), the dissolved oxygen concentration in Chamber 2 (denoted by $S_{O,2}$),  the biodegradable and soluble organic nitrogen concentration in Chamber 2 (denoted by $S_{ND,2}$) and readily biodegradable and soluble substrate concentration in Chamber 1 (denoted by $S_{S,1}$).Units for the respective states are given in Table~\ref{table:table2_wwtp}. The cumulative prediction errors of the chosen states are presented in Table~\ref{table:prediction results}. It can be found that the predictions made by DKOIA are closer to the ground truth compared with the results from DKO, indicating that DKOIA can offer a more accurate model. 

{\color{black}
\begin{rmk}
The modeling results are obtained based on the condition that the known disturbances $p$ to this process, including the flow rate and compositions of the wastewater entering the plant, are available. In the case when the known disturbances are not perfectly known, a practical yet conservative solution would be to average the historical disturbances and use the mean values as a conservative estimate of the current and future values of $p$, which, however, would inevitably lead to degraded modeling and control performance.
\end{rmk}
}

 \subsection{Control performance}

In this section, the proposed iterative Koopman-based MPC method is applied to control the operation of the WWTP. Specifically, the objective is to drive the nitrite nitrogen and nitrate in Chamber 2 (denoted by $S_{NO,2}$) 
and the dissolved oxygen concentration in Chamber 5 ($S_{O,5}$) to a desired set-point.
% The integral of the absolute errors (IAE) of the two states are considered, which are $x_{}$
Since only two states are considered in the MPC problem, with a slight abuse of notation, $x_s = \left[2.1345\, {\rm g\, N \,m^{-3}}, 0.7078\,{\rm g\,(-COD)\,m^{-3}}\right]$, which is the set points of $S_{NO,2}$ and $S_{O,5}$. 
Same as in Section \ref{experiment1}, the controllers based on DKO and NMPC are designed for the WWTP system as baselines.
The weighting matrices of the three controllers are carefully tuned for improved control performance. The weighting matrices $Q$ of the three controllers are $Q_{DKOIA}=\text{diag}(2,1)$, $Q_{DKO}=\text{diag}(1.5,1)$ and $Q_{NMPC}=\text{diag}(1.5,1)$, respectively.
Since no stable inputs $u_s$ can be given under variable known disturbances $p$, matrices $R$ are all set as $R=\text{diag}(0, 0)$.
Considering the challenges of accurately predicting long-term disturbances $p$ in real-world environments, the control horizon is set to 10 for each controller. $l_{max}$ is chosen as 4 for the DKOIA-based controller. 
With the same initial condition, the three controllers are evaluated on the WWTP system under dry weather conditions, with a simulation duration of 14 days.

\begin{figure}[htbp]
  \centering
  \includegraphics[width=1\textwidth]{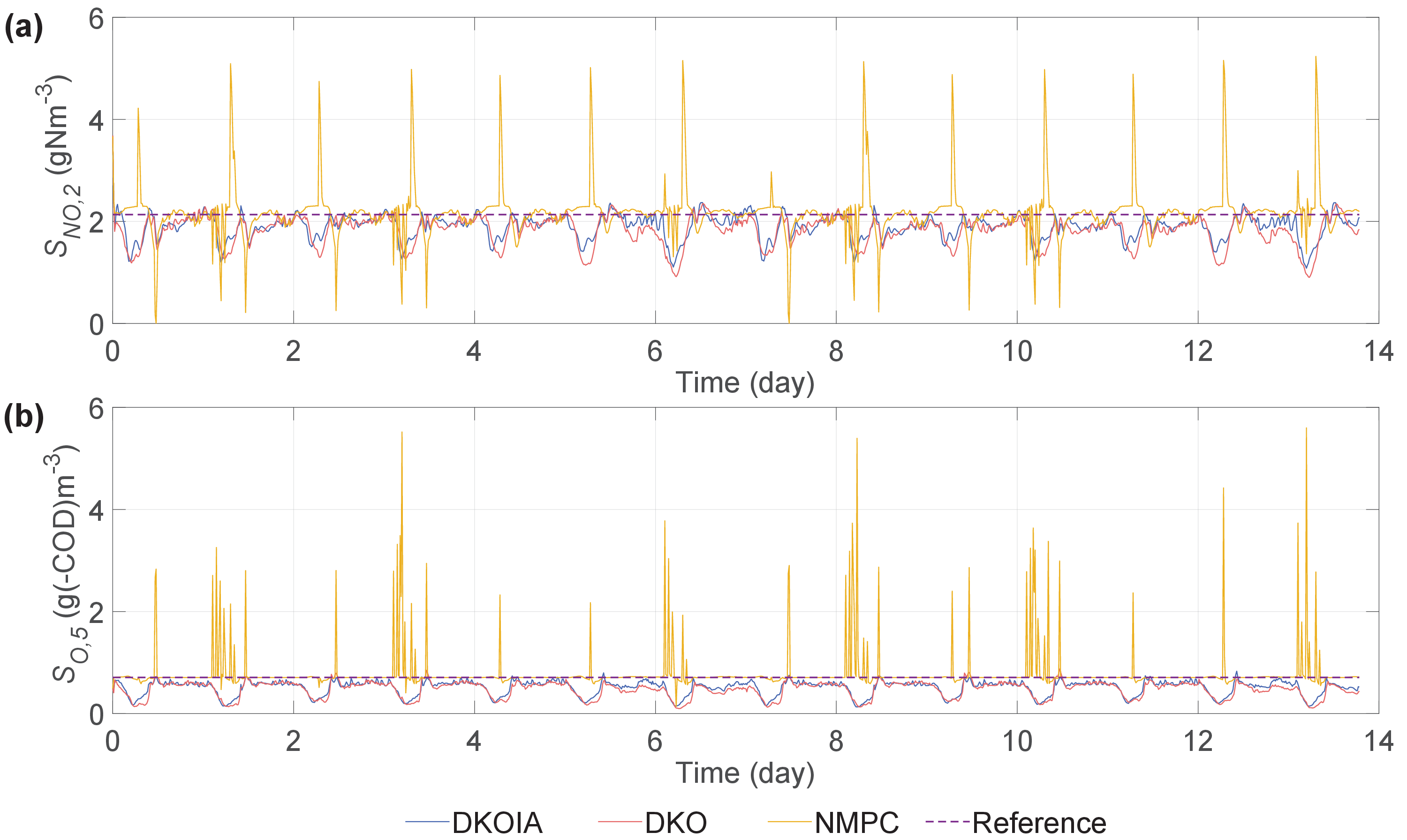}
  \caption{Closed-loop trajectories of the two controlled outputs based on the DKOIA-based controller, the DKO-based controller, and NMPC: (a) $S_{NO,2}$ and (b) $S_{O,5}$.} 
  \label{figure:tracking}
\end{figure}

\begin{figure}[htbp]
  \centering
    \includegraphics[width=1\textwidth]{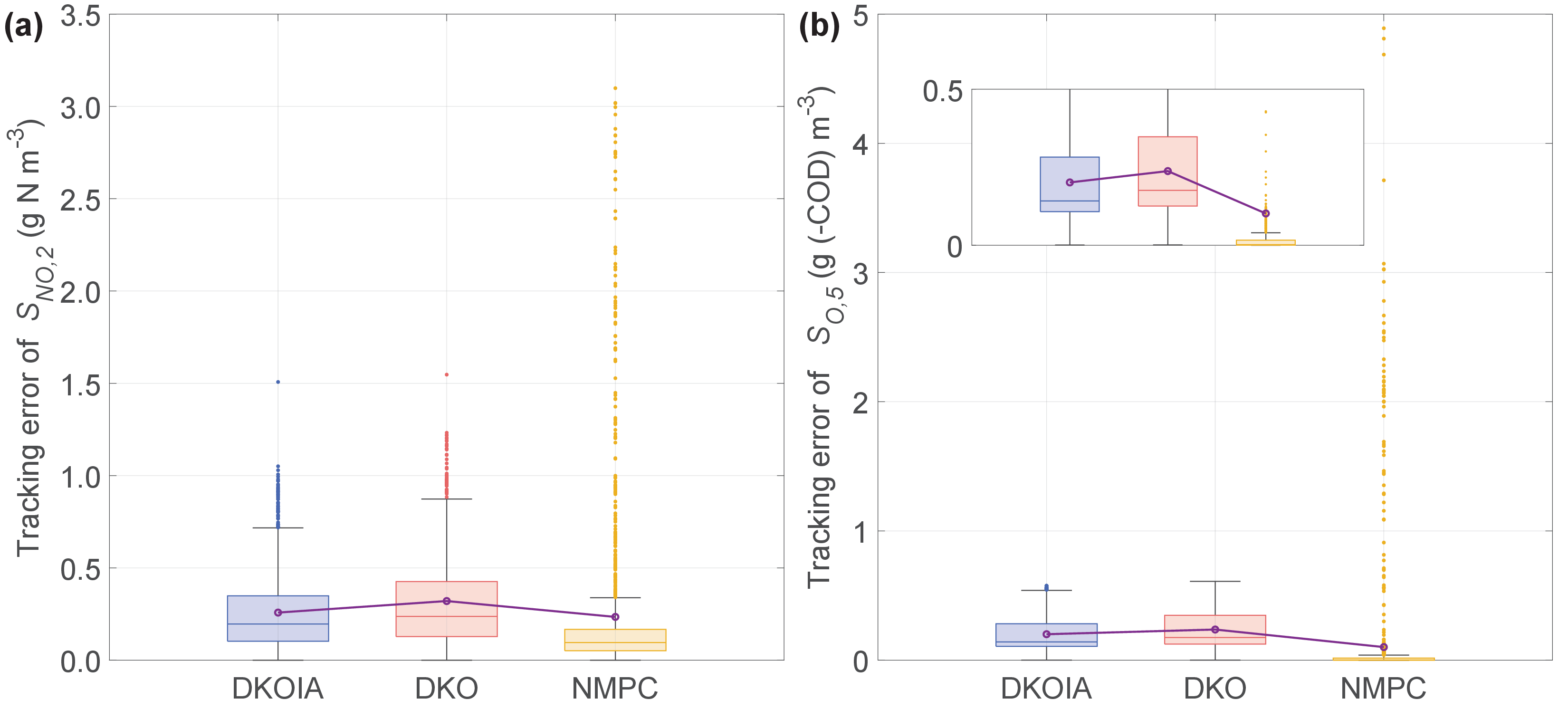}
\caption{Box charts of the tracking errors under the DKOIA-based controller, the DKO-based controller, and NMPC: (a) is for $S_{NO,2}$, and (b) is for $S_{O,5}$.
Purple lines indicate the mean tracking errors of the three controllers. 
} 
  \label{figure:box}
\end{figure}

\begin{figure}[htbp]
  \centering
  \includegraphics[width=1\textwidth]{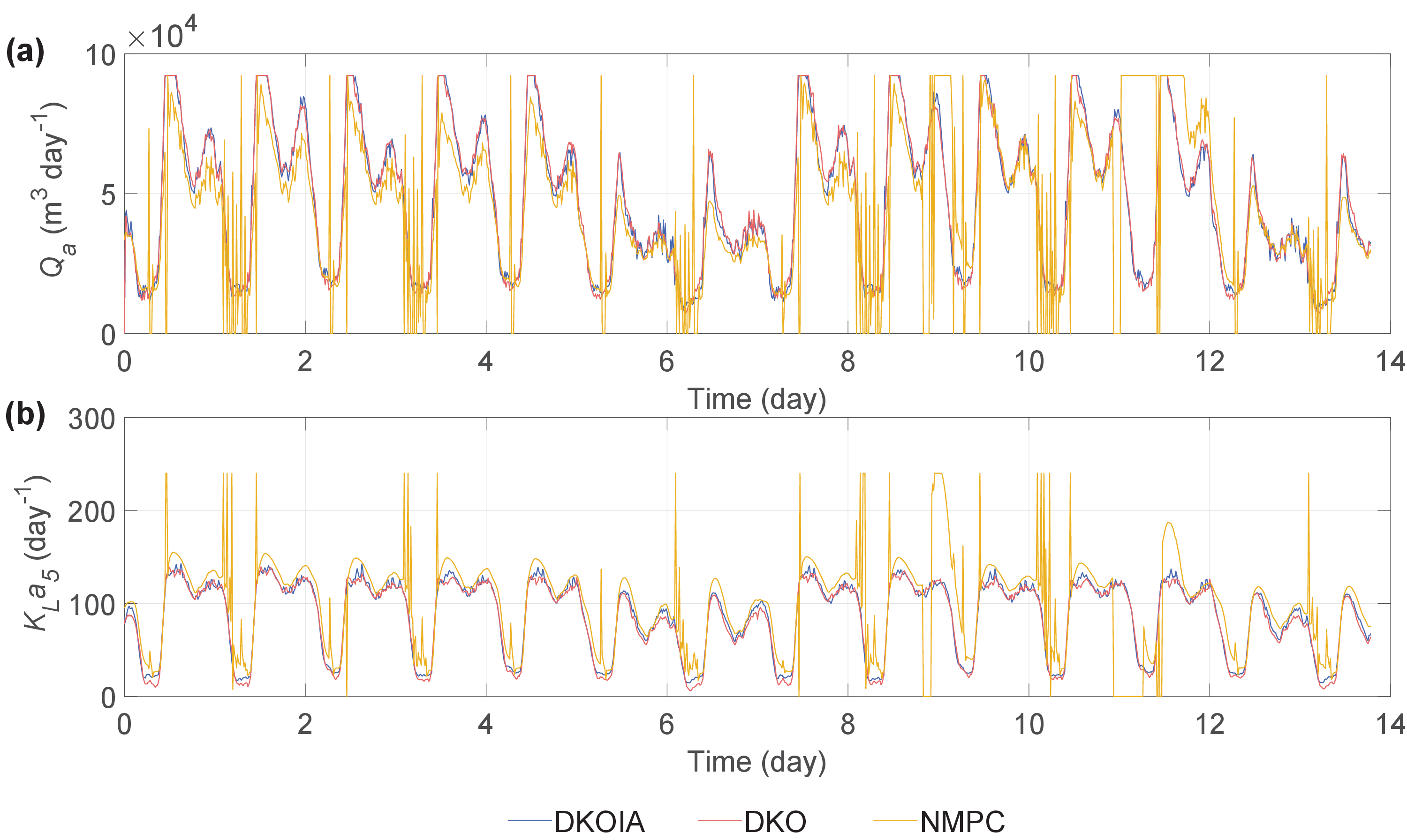}
  \caption{Trajectories of the control inputs based on the DKOIA-based controller, the DKO-based controller, and NMPC: (a) is for $Q_a$, and (b) is for $K_La_5$. } 
  \label{figure:control}
\end{figure}

The trajectories of $S_{NO,2}$ and $S_{O,5}$ are illustrated in Figure~\ref{figure:tracking}.
The system under the DKOIA-based controller achieves a faster convergence to the set point after the large deviation from the disturbance and produces less tracking error when approaching the set point. The tracking errors produced by the three controllers are presented in Figure~\ref{figure:box}. 
The box chart shows the distribution of the tracking errors. 
The mean tracking errors for $S_{NO,2}$ under DKOIA-based controller, DKO-based controller, and NMPC are $0.2851\, {\rm g\, N \,m^{-3}}$, $0.3206\, {\rm g\, N \,m^{-3}}$, $0.2346\, {\rm g\, N \,m^{-3}}$, respectively. The mean tracking error of $S_{O,5}$ under DKOIA-based controller, DKO-based controller, and NMPC are $0.2012  \, \rm g\,(-COD)\,m^{-3}$, $0.2376  \,\rm g\,(-COD)\,m^{-3}$, $0.1018  \,\rm g\,(-COD)\,m^{-3}$, respectively.
Compared with the DKO-based controller, the DKOIA-based controller reduces 19.49$\%$ and 15.29$\%$ tracking errors on $S_{NO,2}$ and $S_{O,5}$, respectively.
Although NMPC exhibits the smallest tracking error, numerous outliers are present in the process, leading to considerable standard deviations. This phenomenon can be attributed to the infeasible solution of the nonlinear MPC.  Control inputs from NMPC display larger fluctuations compared with the other two methods, resulting in relatively large tracking errors under fluctuating inputs, as shown in Figure~\ref{figure:control}. It is worth mentioning that 66 instances of infeasible solutions from NMPC take place throughout the simulated process operation period, while the other two methods using convex optimization consistently find feasible solutions. The proposed DKOIA-based controller can achieve small tracking errors with relatively stable control inputs. 

\section{Conclusion}

In this work, a machine learning-based Koopman modeling approach with input augmentation, which incorporates both states and known inputs into nonlinear mapping was proposed. As compared to the existing learning-based Koopman modeling approaches, the approach proposed in this work holds the promise to provide improved predictive capability, yet it produces Koopman models with nonlinearity in the control inputs. Based on the developed model, a Koopman-based MPC design is formulated. To address the nonlinearity arising from the established Koopman model with input augmentation, an iterative implementation method was proposed, which solves a convex optimization problem iteratively at each sampling instant to approach the optimal solution to the original nonlinear optimization problem associated with the formulated Koopman-based MPC design.
A benchmark chemical process and a biological water treatment process were leveraged to evaluate the performance of the approach on modeling and control. The proposed method provides accurate predictions of the states and good closed-loop control performance for both processes. From the simulation results, the proposed approach largely reduces the modeling error compared with a baseline Koopman modeling approach that does not project known inputs into a higher-dimensional state space. Owing to the improved predictive capability of the model based on the proposed method, the associated controller outperforms the baseline Koopman MPC method in the application of both the chemical process and the water treatment process.

\section*{Acknowledgment}
This research is supported by the Ministry of Education, Singapore, under its Academic Research Fund Tier 1 (RG63/22 \& RS15/21). This research is also supported by the National Research Foundation, Singapore, and PUB, Singapore’s National Water Agency under its RIE2025 Urban Solutions and Sustainability (USS) (Water) Centre of Excellence (CoE) Programme, awarded to Nanyang Environment \& Water Research Institute (NEWRI), Nanyang Technological University, Singapore (NTU).

%%%%% CLEAR DOUBLE PAGE!
\newpage{\pagestyle{empty}\cleardoublepage}

%%%%% CLEAR DOUBLE PAGE!
\newpage{\pagestyle{empty}\cleardoublepage}
\end{document}